\documentclass[twocolumn,aps,amsmath,amssymb,prl]{revtex4}

\usepackage{graphicx}
\usepackage{amsmath}
\usepackage{bm}

\begin{document}

\title{Spin-injection spectra of CoFe/GaAs contacts: dependence on Fe concentration, interface and annealing conditions}

\author{G. Salis}
\email{gsa@zurich.ibm.com}
\author{S. F. Alvarado}
\author{A. Fuhrer}

\affiliation{IBM Research - Zurich,
S\"aumerstrasse 4, 8803 R\"uschlikon, Switzerland}

\begin{abstract}
Spin injection from Co$_{70}$Fe$_{30}$ and Fe contacts into bulk GaAs(001) epilayers is studied experimentally. Using nonlocal measurements, the spin polarization of the differential conductance is determined as a function of the bias voltage applied across the injection interface. The spectra reveal an interface-related minority-spin peak at forward bias and a majority-spin peak at reverse bias, and are very similar, but shifted in energy, for Co$_{70}$Fe$_{30}$ and for Fe contacts. An increase of the spin-injection efficiency and a shift of the spectrum correlate with the Ga-to-As ratio at the interface between CoFe and GaAs.
\end{abstract}

\maketitle

A common ingredient in spin-based device concepts is the generation of spin polarization in a semiconductor by an electrical current drawn from a ferromagnetic contact. Significant spin injection can only be achieved if a spin-dependent contact resistance is introduced~\cite{Rashba2000}. Schottky tunnel-barriers naturally provide such a contact resistance, and spin injection from Fe into GaAs has been demonstrated using both optical~\cite{Zhu2001} and electrical~\cite{Lou2007,Erve2007} detection schemes. The spin polarization in the semiconductor depends in a non-trivial way on the bias voltage $U_c$ applied across the contact resistance, and its relation to the spin-polarized density of states (SPDOS) of the ferromagnetic film is unclear. For Fe, a sign reversal of the spin polarization has been observed that occurs at either reverse or forward bias, depending delicately on the sample growth conditions and on post-growth annealing~\cite{Schultz2009,Salis2010}. Ab-initio calculations~\cite{Butler1997,Freyss2002,Demchenko2006,Chantis2007} suggest that this feature is related to a minority-spin surface state and should be absent for Co$_{70}$Fe$_{30}$ on GaAs~\cite{Chantis2007}. An optical investigation based on the polar Kerr effect~\cite{Kotissek2007} did not find a sign change in samples with Co$_{70}$Fe$_{30}$ contacts.

Here, we study spin-injection spectra in devices consisting of Co$_{70}$Fe$_{30}$ contacts on an n-doped GaAs channel, and compare them with spectra obtained with Fe contacts. To elucidate the role of the interface, the contacts were grown on (001) GaAs surfaces with various surface reconstructions. We employ the non-local measurement technique~\cite{Johnson1985} to determine the spin polarization. In the reverse-bias direction, the polarization of the injected electrons is found to be of majority spin, whereas in the forward direction we find a sign reversal at $U_c\approx-100$\,mV, similar to the Fe/GaAs system. We analyze the data in terms of the differential conductance $P_\Sigma$ of the injection contact. Its dependence on $U_c$ reveals spectral information on the SPDOS. For all Co$_{70}$Fe$_{30}$ samples and annealing conditions, we observe a minority-spin peak at $U_c\approx-100$\,mV and a majority-spin peak at $\approx100$\,mV. This spectrum is very similar to that found for Fe contacts, but shifted by about 30\,mV towards larger $U_c$, which is consistent in sign --- but not in magnitude --- with the higher band filling of bulk Co$_{70}$Fe$_{30}$. Together with the observed dependence of $P_\Sigma$ on the measurement temperature $T$ for $T<50$\,K, the interface preparation and the post-growth annealing, we conclude that the spectra can be explained by the superposition of a minority-spin-related interface state with majority-spin injection from the bulk of the ferromagnet.

A Si-doped (5$\times$10$^{16}$\,cm$^{-3}$) GaAs epilayer with a 15-nm-thick highly-doped Schottky tunneling barrier and a surface protected by an As cap was grown on a semi-insulating GaAs(001) wafer using molecular-beam epitaxy, see Ref.~\onlinecite{Salis2009} for details. The substrates were transferred into another vacuum chamber, where the As cap layer was removed by heating at temperatures $T_i$ of 400, 450 or 500$^\circ$C for 1 h. For $T_i=400$\,$^\circ$C, scanning tunneling microscopy confirmed an As-rich c(4$\times$4) reconstruction. At higher As desorption temperatures, the surface stoichiometry of the substrate changes from As- to Ga-rich, as indicated by changes in the surface reconstruction. The substrates were then cooled to -50\,$^\circ$C, and 5\,nm of Co$_{70}$Fe$_{30}$ or Fe was grown by sublimation and capped by a 2-nm-thick Au layer. Using optical lithography and Ar sputtering, bar-shaped ferromagnetic contacts were defined and contacted with a Ti/Au layer. Measurements of $\Delta U_{nl}$ were done after each of the 10-minute-long post-growth annealing steps at $T_a=120$, 170, 200, 230 and 260$^\circ$C.

\begin{figure}[ht]
\includegraphics[width=70mm]{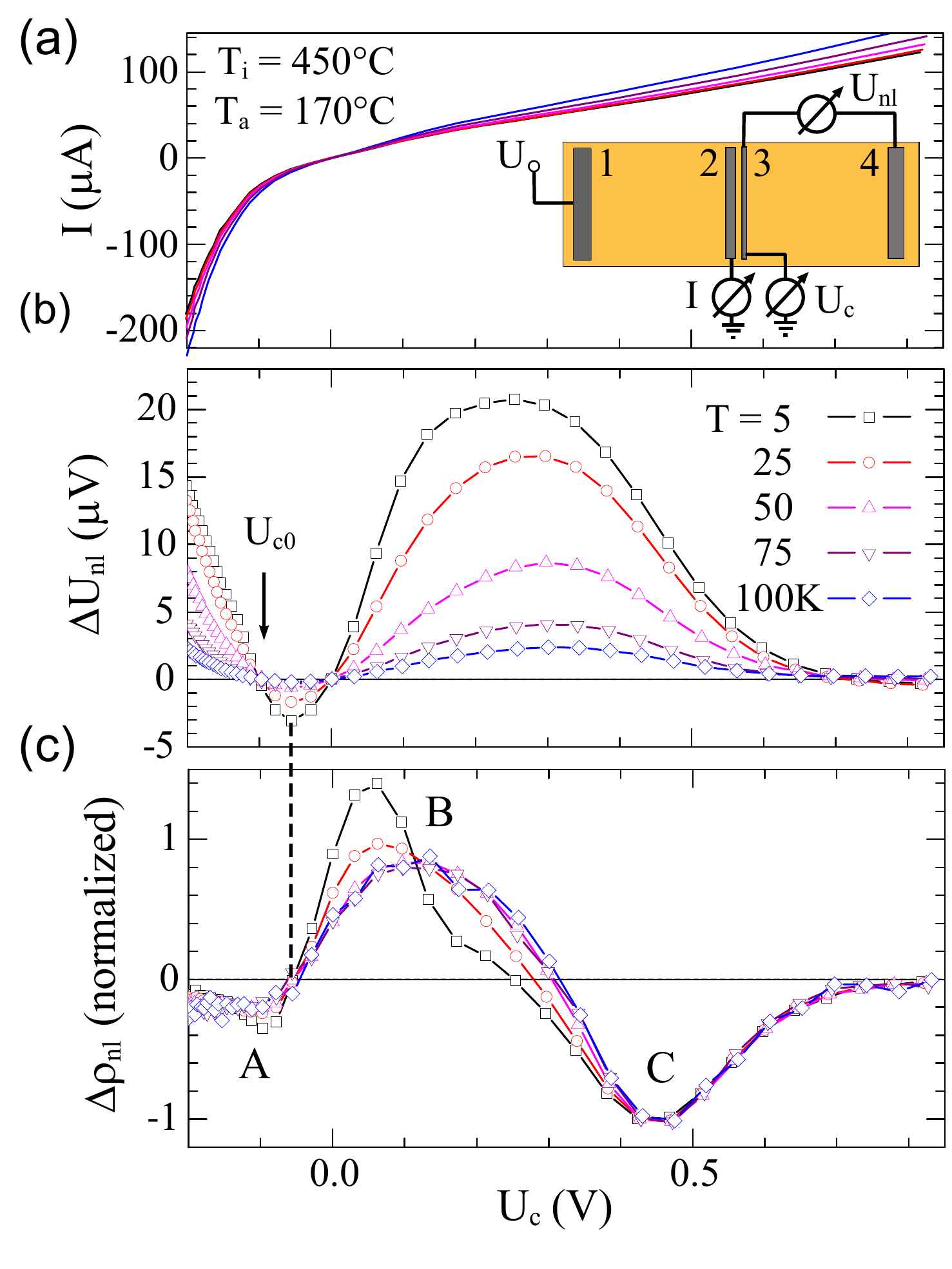}
\caption{\label{fig:fig1} (Color online) Data on Co$_{70}$Fe$_{30}$/GaAs sample for $T_a=170^\circ$C and $T_i=450^\circ$C. (a) $I$ vs. $U_c$ for different $T$; inset shows measurement geometry. (b) The measured nonlocal voltage $\Delta U_{nl}$ is positive for $U_c>0$ (majority-spin injection), changes sign at $U_c=0$, and becomes positive again for $U_c<-100$\,mV. (c) Peak A (B) of minority (majority) spin injection become visible in $\Delta\rho_{nl}=d\Delta U_{nl}/dI$. Peak C at $\approx 450$\,mV is related to a decrease of the spin polarization at high reverse bias.}
\end{figure}

In each sample, four 60-$\mu$m-long ferromagnetic bars are aligned along the GaAs [110] direction, which is their magnetic easy axis~\cite{Bianco2008}. The two outer bars labeled 1 and 4 in the inset of Fig.~\ref{fig:fig1}(a) are 10\,$\mu$m wide and serve as reference contacts. A current $I$ between contact 1 and the 6-$\mu$m-wide injection contact 2 generates spin polarization in the GaAs channel layer below contact 2. A voltage $U_{nl}$ between the 2-$\mu$m-wide detection contact 3 and contact 4 was measured for antiparallel and parallel magnetization of contacts 2 and 3 as set by an external magnetic field applied along the [110] direction. The difference $\Delta U_{nl}$ between these two magnetic configurations is proportional to the spin polarization below contact 3 that has diffused from contact 2 across the 3-$\mu$m-wide gap.

Measurements on a sample with $T_i=450$\,$^\circ$C and $T_a=170$\,$^\circ$C for different $T$ are summarized in Fig.~\ref{fig:fig1}. Irrespective of $T$, significant current flows in reverse bias [Fig.~\ref{fig:fig1}(a)], indicating that tunneling dominates the transport across the Schottky barrier. In Fig.~\ref{fig:fig1}(b), $\Delta U_{nl}(U_c)$ is plotted. The positive $\Delta U_{nl}$ for $U_c>0$ corresponds to injection of majority spins. $\Delta U_{nl}$ crosses zero at $U_c=U_{c0}\approx -100$\,mV, and becomes positive again for $U_c<U_{c0}$, marking a transition where predominantly minority spins flow into the Co$_{70}$Fe$_{30}$ layer, leaving behind majority spins in the GaAs. This is similar to what is found in samples with Fe contacts and is attributed to the opening of a minority-spin channel for forward-bias conditions~\cite{Lou2007,Salis2009}.

In Fig.~\ref{fig:fig1}(c), we analyze the data by plotting $\Delta\rho_{nl}=d\Delta U_{nl}/dI$. $\Delta U_{nl}$ is proportional to the product of $I$ and the spin polarization of the injection current, thus $\Delta\rho_{nl}$ is proportional to the spin polarization $P_{\Sigma}=(\Sigma_+-\Sigma_-)/(\Sigma_++\Sigma_-)$ of the differential conductance $\Sigma_{\pm}=dI_{\pm}/dU_c$ of the injection contact~\cite{Valenzuela2005}, where the subscript $\pm$ refers to the relative alignment of the spins with the magnetization of the detection contact. In tunneling spectroscopy, $\Sigma$ versus $U_c$ energetically resolves the density of states involved. Similarly, $P_{\Sigma}$ probes the SPDOS of the injection contact. However, the relation between $P_{\Sigma}$ and the SPDOS is complicated by the bias-induced change of the Schottky tunnel barrier, affecting the overall as well as the relative transmission probability of $sp$- and $d$-derived states~\cite{Alvarado1995}. In addition, the summation over the in-plane momenta, the band structure of GaAs and the contribution of interface-related states need to be taken into account. Close to zero bias, we find two distinct peaks in $\Delta\rho_{nl}$ labeled A (minority spin) at about -100\,mV and B (majority spins) at $\approx$100\,mV [see Fig.~\ref{fig:fig1}(c)]. At $U_c\approx -60$\,mV, $\Delta\rho_{nl}$ crosses zero, and $\Delta U_{nl}$ exhibits a negative peak [dashed line in Fig.~\ref{fig:fig1}(c)], marking the situation where minority and majority spins have the same differential conductance. For $U_c<-60$\,mV, more and more majority spins are left behind in the GaAs layer, until at $U_c=U_{c0}$, $\Delta U_{nl}=0$. Note that $U_{c0}$ can be found by solving $\int_0^{U_{c0}} \Delta\rho_{nl}dI=0$.

\begin{figure}[ht]
\includegraphics[width=70mm]{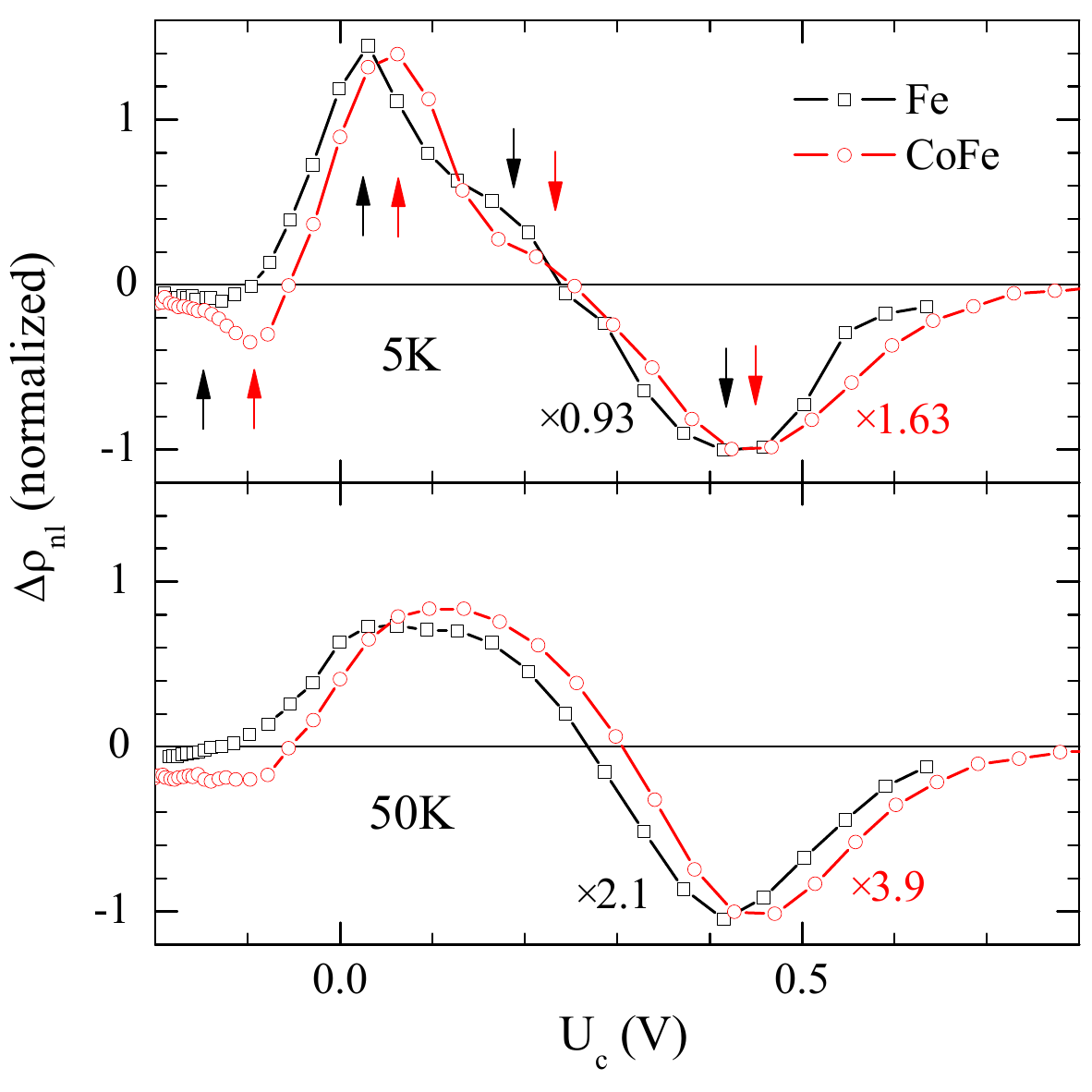}
\caption{\label{fig:fig2} (Color online) Comparison of Fe and Co$_{70}$Fe$_{30}$ contacts: $\Delta\rho_{nl}(U_c)$ normalized to peak C at $T=5$ and 50\,K. A shift of all peaks in the Co$_{70}$Fe$_{30}$ spectrum (arrows) towards higher $U_c$ is observed, compatible with a higher band filling of Co$_{70}$Fe$_{30}$. Data collected for samples with $T_i=450^\circ$C and $T_a=230^\circ$C (Co$_{70}$Fe$_{30}$), and $T_i=400^\circ$C and $T_a=170^\circ$C (Fe).}
\end{figure}

For Fe, a strong and sharp peak in the minority-spin SPDOS related to interface states~\cite{Butler1997,Freyss2002,Demchenko2006,Chantis2007} is expected. Measured $d_{3z^2-r^2}$-orbital-derived surface states of bcc(001) Fe and Cr~\cite{Stroscio1995} are centered at similar energies, suggesting that the interface-related state of Co$_{70}$Fe$_{30}$/GaAs is not much shifted relative to that of Fe/GaAs. In addition to this minority-spin channel, the majority-spin band of bulk $\Delta_1$ states gives rise to broad majority-spin injection up to $U_c=1$\,V~\cite{Wunnicke2002}. An interplay of these two channels accounts for the observation of peaks A and B. The disappearance of $\Delta U_{nl}$ at large positive $U_c$ leads to a negative peak C in $\Delta \rho_{nl}$ at $U_c\approx$~450\,mV. In all our measurements, $\Delta U_{nl}$ approaches zero at large reverse bias. Therefore, we attribute this peak to a decrease in spin-injection efficiency rather than an opening of a new minority-spin channel. At $U_c>300$\,mV, increasing transmission into the L and X valleys of GaAs could reduce the spin polarization, in addition to the spin relaxation at the $\Gamma$-point induced by spin-orbit interaction~\cite{Saikin2006,Song2010}.

\begin{figure}[ht]
\includegraphics[width=75mm]{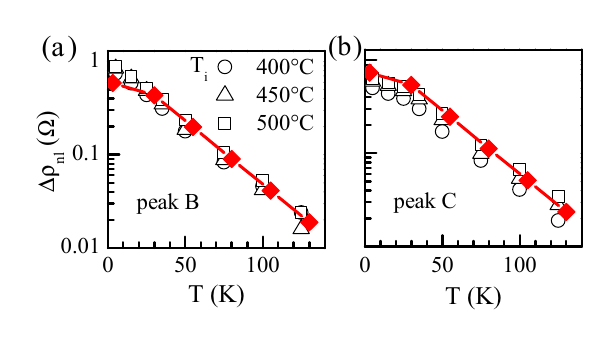}
\caption{\label{fig:fig3} (Color online) $\Delta\rho_{nl}$ of (a) peak B and (b) peak C versus $T$ for samples with different $T_i$. The solid (red) lines and (red) diamonds indicate the temperature dependence attributed to the spin decay in the GaAs channel, as determined separately from Hanle curves. A small increase of the peak height with $T_i$ is observed.}
\end{figure}

Co$_{70}$Fe$_{30}$ has a bulk band structure that is similar to that of Fe, but the bands are filled to a higher level~\cite{Turek1994}. If peaks A and B were related to the band-structure of the ferromagnetic metal, one would expect $\Delta\rho_{nl}(U_c)$ to shift towards lower $U_c$ when replacing Co$_{70}$Fe$_{30}$ with Fe. The measured $\Delta\rho_{nl}(U_c)$ of a sample with Fe contacts is about 1.8 times larger, but otherwise surprisingly similar to that of samples with Co$_{70}$Fe$_{30}$ contacts, including a splitting of the central majority peak at $T=5$\,K (see Fig.~\ref{fig:fig2}). The entire Fe spectrum is shifted in $U_c$ by about $-30$\,mV compared with the Co$_{70}$Fe$_{30}$ spectrum, with a less pronounced and slightly more strongly shifted peak A. The expected change in the band filling of the bulk band structure is, however, much larger than 30\,meV, by about one order of magnitude, and moreover distributed differently among the minority and majority bands because of the smaller exchange splitting of Co$_{70}$Fe$_{30}$~\cite{Schwartz1994,Turek1994}. This demonstrates that the specific shape of the spectra cannot stem from the bulk band-structure alone, in agreement with assigning peak A to an interface-related minority-spin state. However, it is surprising that peak C, which we attribute to a loss of spin polarization at large reverse bias, also shifts when changing the ferromagnetic material.

The data in Fig.~\ref{fig:fig1}(c) are normalized to the height of peak C, whose shape remains remarkably constant with $T$. With decreasing $T<50$\,K,  peak B separates into a sharp peak of increasing height at $U_c=50$\,mV, and a broader peak centered at 200\,mV. Figures~\ref{fig:fig3}(a) and (b) show the height of peaks B and C versus $T$ for Co$_{70}$Fe$_{30}$ samples, prepared at different $T_i$ and annealed at $T_a=230^\circ$C. Peak heights were taken as the maximum positive and negative values of $\Delta\rho_{nl}$. They strongly decrease with $T$, mostly because of the spin decay in the GaAs channel, as estimated from the diffusion constant and spin lifetime obtained from Hanle measurements~\cite{Salis2010} at different $T$. We plot this contribution of the spin decay in Fig.~\ref{fig:fig3} as red (solid) diamonds connected by a line. One data point at 3\,K was obtained in a different sample by suppressing dynamic nuclear polarization~\cite{Fuhrer2011}. Above 35\,K, $\Delta\rho_{nl}$ at both peaks follows this line, suggesting that the signal decrease is attributed to the GaAs channel. At lower $T$, however, we find that peak B is enhanced, i.e., that there $P_\Sigma$ decreases with $T$. We note that the bulk band structure and magnetization of the Co$_{70}$Fe$_{30}$ contacts are not expected to change much with $T$. The position $U_c\approx200$\,mV of the broad peak at low $T$, however, coincides with the value where band-profile calculations show a depletion of confined states in the highly n-doped GaAs region. This supports the prediction~\cite{Dery2007} that such states may influence the spin-injection efficiency.

\begin{figure}[ht]
\includegraphics[width=80mm]{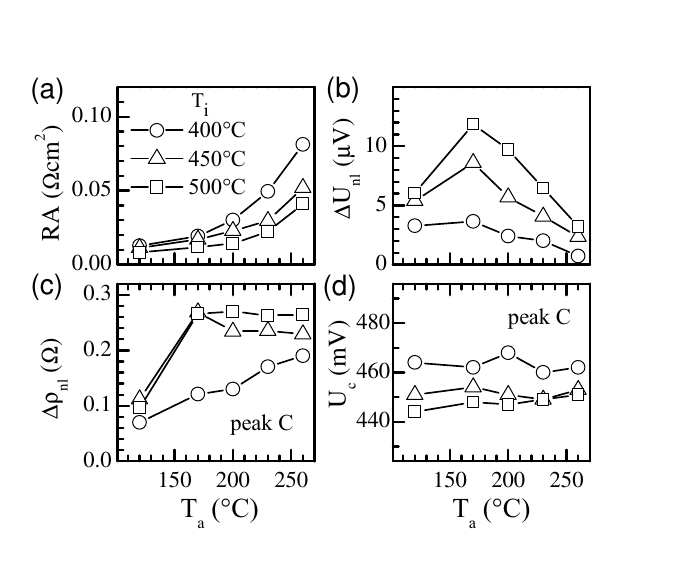}
\caption{\label{fig:fig4} Influence of $T_a$, measured at $T=50$\,K for Co$_{70}$Fe$_{30}$ contacts. (a) $RA$ of the injection contact monotonically increases with $T_a$. (b) The spin polarization (peak value of $\Delta U_{nl}$) first increases at $T_a=170$\,K, and then decreases because less current flows. (c) $\Delta \rho$ measured at peak B increases or remains constant above $T_a=170$\,K. (d) The position $U_c$ of peak B does not depend on $T_a$ and slightly shifts with $T_i$.}
\end{figure}

Post-growth annealing modifies the magnetic behavior of the contact layer~\cite{Bianco2008}, its interface with GaAs~\cite{Zega2006,Shaw2007,Fleet2010}, and the spin-injection efficiency ~\cite{Wang2005,Adelmann2005B,Schultz2009,Salis2010}. In the following, we will first discuss the changes in the resistance area product $RA$ of Co$_{70}$Fe$_{30}$ contacts as obtained from the low-bias slope of $I$ vs. $U_c$, and then present the influence on $\Delta\rho_{nl}$.

$RA$ increases monotonically with increasing $T_a$, see Fig.~\ref{fig:fig4}(a). A similar behavior was observed in Fe/GaAs~\cite{Fleet2010} and Fe/Ga$_{0.9}$Al$_{0.1}$As~\cite{Adelmann2005B}. The latter report proposes an interpretation in terms of the formation of an interfacial ferromagnetic Fe$_3$Ga$_{2-x}$As$_x$ ternary alloy at higher annealing temperatures, whereas the former associates the effect with a sharpening of the Fe/GaAs interface~\cite{Zega2006,Fleet2010}. For increasing $T_i$, $RA$ decreases. In view of the reduced amount of As on the reconstructed surface, this result is not compatible with calculations that predict a higher barrier height on Ga- than on As-terminated GaAs interfaces~\cite{Fleet2010, Demchenko2006,Freyss2002}. On the other hand, it has been shown that intermixing at the interface is expected to strongly reduce the barrier height~\cite{Demchenko2006}, suggesting that a Ga-rich GaAs (001) surface promotes more intermixing than the As-terminated surface.

At fixed $U_c$ and $P_\Sigma$, $I$ and therefore $\Delta U_{nl}$ decrease when $RA$ increases. This is seen in Fig.~\ref{fig:fig4}(b) showing the maximum positive value of $\Delta U_{nl}$ that increases abruptly after the first annealing step at $T_a=170^\circ$C, but then decreases with each subsequent step. However, $\Delta\rho_{nl}$ remains approximately constant after a steep increase at $T_a=170^\circ$C for both $T_i=450$ and 500$^\circ$C [Fig.~\ref{fig:fig4}(c)]. Only for $T_i=400^\circ$C does $\Delta\rho_{nl}$ steadily increase with $T_a$. This suggests that post-growth annealing modifies the metal-semiconductor interface obtained from the As-rich GaAs surface with c(4$\times$4) reconstruction more strongly than that obtained at higher $T_i$. Considering that annealing promotes the diffusion of As from the interface through the Co$_{70}$Fe$_{30}$ layer~\cite{Bianco2008}, the increase in $\Delta\rho_{nl}$ with $T_a$ is an indication of increased spin-injection efficiency for interfaces that contain less As. This is compatible with the observed weak increase of $\Delta\rho_{nl}$ with $T_i$ [see Figs.~\ref{fig:fig3} and \ref{fig:fig4}(c)]. The sharp increase of $\Delta\rho_{nl}$ upon annealing at $T_a=170^\circ$C indicates a mechanism relating the structural properties of the interface with the GaAs surface stoichiometry. The shape of $\Delta\rho_{nl}$ versus $U_c$ does not change significantly with annealing. As shown in Fig.~\ref{fig:fig4}(d), the position of peak C depends only weakly on $T_a$, but clearly shifts towards smaller $U_c$ with increasing $T_i$. This behavior affects the entire spectrum and is not restricted to peak C. The shift towards lower $U_c$ correlates with a decrease in $RA$, which is also observed when replacing Co$_{70}$Fe$_{30}$ with Fe, but not when annealing the samples. For Fe ($T_i=400^\circ$C, $T_a=170^\circ$C), we measure $RA=3.6\times10^{-3}$\,$\Omega$cm$^2$, about one order of magnitude less than for Co$_{70}$Fe$_{30}$ and in agreement with the smaller work function and thus smaller Schottky-barrier height of Fe~\cite{Turek1994}. We suspect a lower $d$-band filling at the interface to be responsible for the observed shift in the spectra for both  increasing Ga interface concentration predicted in Ref.~\onlinecite{Demchenko2006}, and increasing Fe concentration.

In conclusion, we have investigated spin injection from Co$_{70}$Fe$_{30}$ and Fe contacts into GaAs. Spectra of $\Delta\rho_{nl}$ versus $U_c$ are related to the spin polarization of the differential conductivity $P_\Sigma$ of the tunnel barrier. The spectra of Fe contacts are remarkably similar to those of Co$_{70}$Fe$_{30}$ contacts, but shifted by about 30\,mV towards smaller $U_c$. A shift into the same direction, but smaller, is observed for Co$_{70}$Fe$_{30}$ grown on a Ga-rich as compared with an As-rich GaAs surface. For Co$_{70}$Fe$_{30}$, we find a sharp increase of $P_\Sigma$ with a first annealing step at $T_a=170^\circ$C. For higher $T_a$, $P_\Sigma$ further increases only for an As-rich GaAs surface. For both Co$_{70}$Fe$_{30}$ and Fe contacts, the spin-injection spectra can be understood as the combination of a surface-state-related minority-spin channel with a broader bulk majority-spin channel. A strong modification with $T<50$\,K of the majority-spin peak at reverse bias suggests that depletion of confined states in the heavily-doped GaAs surface region also influences the spectra. These results are important for further optimizing the spin-injection efficiency and suggest to exploit the interplay of surface states with bulk states to determine efficiencies in Fe and Co$_{70}$Fe$_{30}$ contacts on GaAs.

We thank R. Allenspach, I. Neumann and S. O. Valenzuela for fruitful discussions and D. Caimi, U. Drechsler, M. Tschudy and M. Witzig for technical support.


\end{document}